Short Paper*

# Data Analysis of Bulacan State University Faculty Scientific Publication Based on Google Scholar using Web Data Scraping Technique

Jayson M. Victoriano
Information and Industrial Technology Department
Bulacan State University-Sarmiento Campus, Philippines
jayson.victoriano@bulsu.edu.ph
(corresponding author)

Jaime P. Pulumbarit
College of Information and Communication Technology
Bulacan State University-Malolos Campus, Philippines
jaime.pulumbarit@bulsu.edu.ph

Luisito Lolong Lacatan
College of Engineering, Laguna University, Philippines
louie1428@gmail.com

Richard Albert S. Salivio
Information and Industrial Technology Department
Bulacan State University-Sarmiento Campus, Philippines
richard.salivio@bulsu.edu.ph

Rica Louise A. Barawid
General Academic and Teacher Education Department
Bulacan State University-Sarmiento Campus, Philippines
ricalouise.barawid@bulsu.edu.ph

## Abstract

*Purpose* – The paper aims to analyze and monitor the research publication productivity of the faculty members of Bulacan State University. This paper compiles all the scientific publications from Bulacan State University (BulSU) and its external campuses as an index in Google Scholar. This study was intended to track and monitor the scientific productivity of the faculty members of Bulacan State University and each college and campus.

*Method* – With the use of web data scraping techniques, metadata files were gathered from various faculty research engagement in Google Scholar Database. Web scraping is the process of extracting useful information from web pages (HTML pages) and saving it in a format that you specify, such as Excel (.xls), CSV (comma-separated values), or any other structured format using PurseHub. Different stages of data extraction are completed before descriptive insight is provided to the university. The researchers used the web data scraping techniques to extract the necessary data in the web through the Google Scholar database to specifically gather the research outputs of the faculty members of Bulacan State University.

*Results* – Results have shown that a significant increase that happened throughout the covered years on the number of publications was made possible by the scholarships and study breaks that the faculty members were able to get and a significant increase in the number of citations was brought by the benefits of collaborating with other researchers from other HEIs.

*Conclusion* – Bulacan State University has seen both big improvements and stagnation with the growth of the number of publications being produced by the faculty members. Based on the results of the study, the publications of BulSU have experienced a significant increase due to the impact of scholarship and the benefits that comes with it to the productivity of the faculty members. On the other hand, the factor that affected the significant increase in citations is the collaborators with researchers from other academic institutions.

*Recommendation* – The administration should consider doing necessary steps to encourage the faculty members who engage in research activities to create Google Scholar accounts for easy monitoring of publications and citations. This would allow the easier navigation of the research studies produced by the faculty members which will help the university monitor the progress of research activity effectively and for an impressive research activity, the researchers also recommend that the administration should also provide competitive incentives and positive reinforcements. The faculty, on the other hand, are recommended to broaden their horizon and engage in collaborations with other researchers from other academic institutions for wider perspective of things.

*Practical Implication* – Maximizing the use of data scraping in gathering information about research activity from vast networks will make the navigation and organization of research activity easier. With this, monitoring the progress of faculty research productivity will also be done in ease.

## INTRODUCTION

A university is a professional organization (van Rijnsoever, 2008), for which success depends to a large extent on the work of its researchers with this the number of indexed, peer-reviewed journal papers published by an individual or institution is a well-known metric for research and development (R&D) productivity (Vinkler, 2010), Bulacan State University (BulSU) as a premier university in the region and the Philippines aim to accelerate research output through different means and discipline, and part of the BulSU long term plan to be a research university in the future. Research Management Office of the University used a different approach to motivate and uplift research culture in the University, Using Google Scholar as a freely accessible web search engine that indexes the full text or metadata of scholarly literature across an array of publishing formats and disciplines, this will be the basis for the university to check and monitor the research output of the faculty and staff of the university community (David, 2019; Saddler, 2010).

Since the data are already available on the internet using Google Scholar, Google Scholar is a huge library of scientific literature that lets users search for material, cross-reference it with other sources, and stay up to date on new research as it is published. With this, to extract data online different techniques are used and utilized (Cloudflare, 2020). data as and form the basis of the study and evaluation. Using data scraping tools such as Data Scraper developed by Google, PurseHub, Content Grabber, Web Content Extractor, and Scrapy (Keval, 2015; Saurkar, 2018) to name a few, this tool used a process of transferring data from a website to a spreadsheet or a local file to our PC. It is one of the fastest ways to collect data from the web and, in some situations, to send that data to another website.

The paper aims to visualize the research engagement output of BulSU faculty members using data analytic tools such as Pursehub. With the help of the visualization, the paper aims to use data scraping techniques in utilizing the Google Scholar database to extract the needed data to be analyzed for recommendation and policy. After achieving the objectives and producing results, the paper seeks to be published in Scopus-index and another high-impact journal.

## METHODOLOGY

Web scraping is a fantastic way to retrieve information from the internet. Data from websites that are unstructured and transforms that information into organized information that can be stored and retrieved in a database, the data is analyzed (Fernandez-Villamor, 2011). Scraping the web is also known as the extraction of data from the internet (Sirisuriya, 2015). Figure 1 demonstrates the Web Scraping process, which includes identifying the website target, collecting the URL of the page, making a request to the URL to acquire the HTML of the page, using locators to find the data in the HTML, and saving the data in JSON, CSV, or another structured data format.

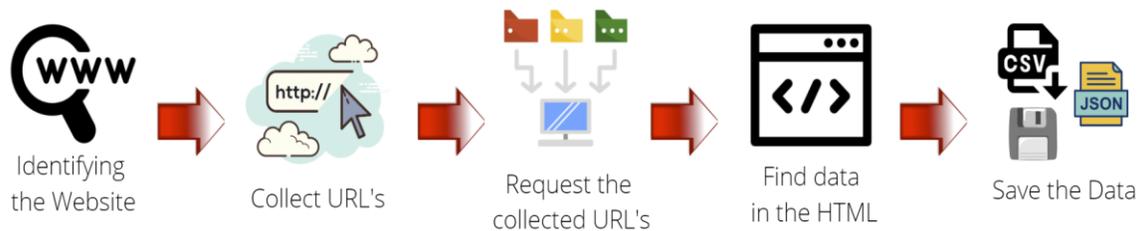

*Figure 1.* Web Scraping process

The researchers employed the Hirsch index (h-index) to compare the effect of published works, in addition to normal publication counts and metrics, to estimate the scientific production of BulSU Faculty. From 2010 to 2021, only journal papers written by teachers and researchers affiliated with BulSU were considered.

Data on individual faculty counts and characteristics came from the updated Google Scholar database. Google Scholar databases were used to count citations for each work which is used to calculate the h-index (Majeed, 2019). Gray literature, unpublished papers, periodicals, and questionable journal titles, books were not used by the researchers because it overestimates citation counts and over-represents citations (Fatima, 2020), the only paper that is published under reputable journal publications and indexes in Scopus, web of science and other are being included in the study.

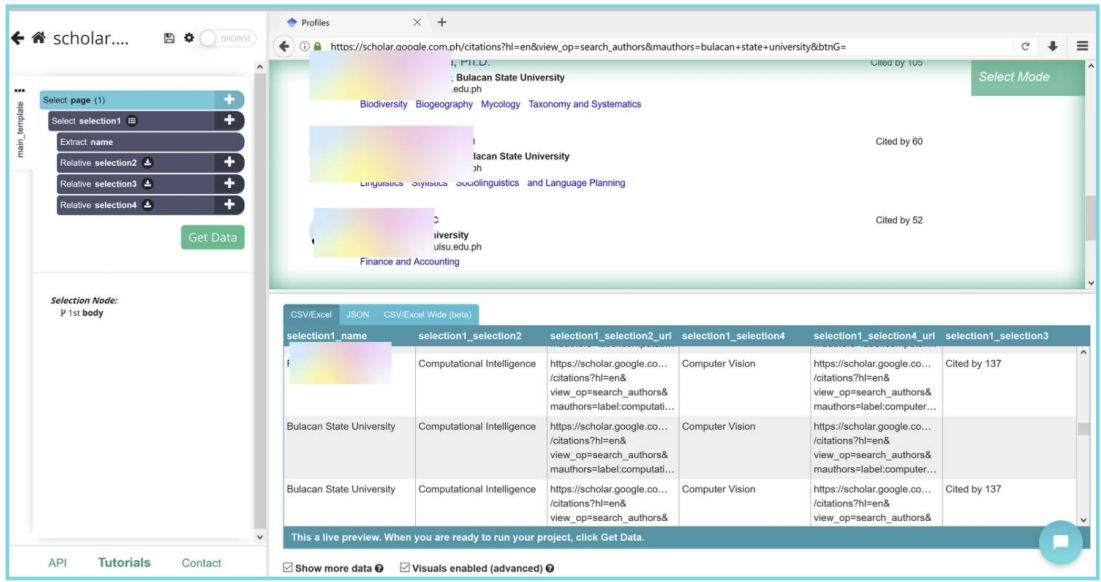

Figure 2 PurseHub Environment

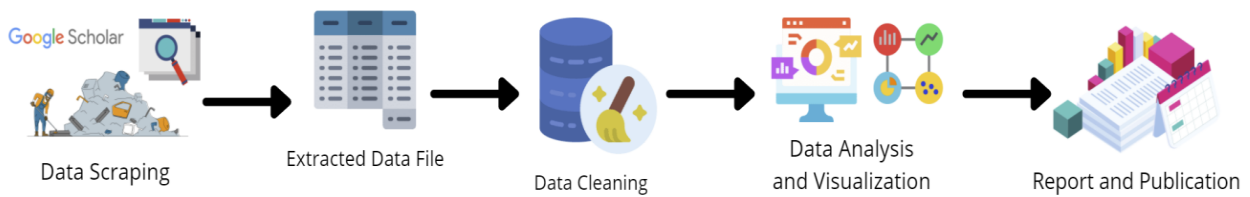

*Figure 3.* Conceptual Framework

The process that this study followed started from the Data Scraping stage shown in Figure 3 where the application, PurseHub has shown in Figure 2, will extract the wanted data from Google Scholar. After the extraction, Data Cleaning will take place wherein unwanted data will be removed so that the only data to be analyzed. Once the data cleaning is done and only the necessary data remains, the data analysis and visualization will begin which will lead to the presentation of the report and then publication.

To set the limitations and scope of the study, the researchers chose to use the web data scraping application to sort and find the publications of BulSU Faculty members on the Google Scholar database. The featured selections shown in Table 1, the different aspects and concepts that the paper focused on such as the year the paper was published, the names of the researchers, the college that the researcher is under, the number of publications under the name of the researcher, the number of citations a particular college or campus received and lastly the collaborations that the researchers were able to be part of.

Table 1. Featured Selections of the paper

| Feature | Type | Description |
|---|---|---|
| Year Published | integer | This featured selection includes the year when the paper was successfully published. |
| Researcher | Text | The name of the faculty member of BulSU that conducted research that was successfully published and is posted in Google Scholar |
| College | Text | The college or campus that the researcher is a part of |
| No of Pub | Numeric | This shows the total number of publications gathered in the researcher's Google Scholar profile. |
| No Citation | Numeric | This is the number of times the paper was cited in other researches. |
| Collab | Text | This featured selection shows the names of the collaborators that worked with the researcher. |

## RESULT

After undergoing a series of steps from the data scraping stage to data analysis and visualization, the Web Data Scraping application was able to gather a total of 179 publications from seventy-five (75) faculty researches out of eleven (11) different colleges and campuses of Bulacan State University found in Google Scholar from the covered years, 2010 – 2021. Different graphs and charts were provided for the visualization of the results.

The researchers have compiled the gathered research outputs of BulSU faculty members from the year 2010 to 2020 shown in Figure 4 Out of all the colleges and campuses of Bulacan State University, the College of Education (COEd) has shown the greatest number of publications from the covered decade with the College of Science (CS) following its lead and just a few publications behind. The next few colleges and campuses have successfully published papers in half the number of the two leading colleges. As we look closely at the red line representing Sarmiento Campus, it showed a significant increase in number from having no publications from the year 2010 to 2017 to producing sixteen (16) publications in just three years.

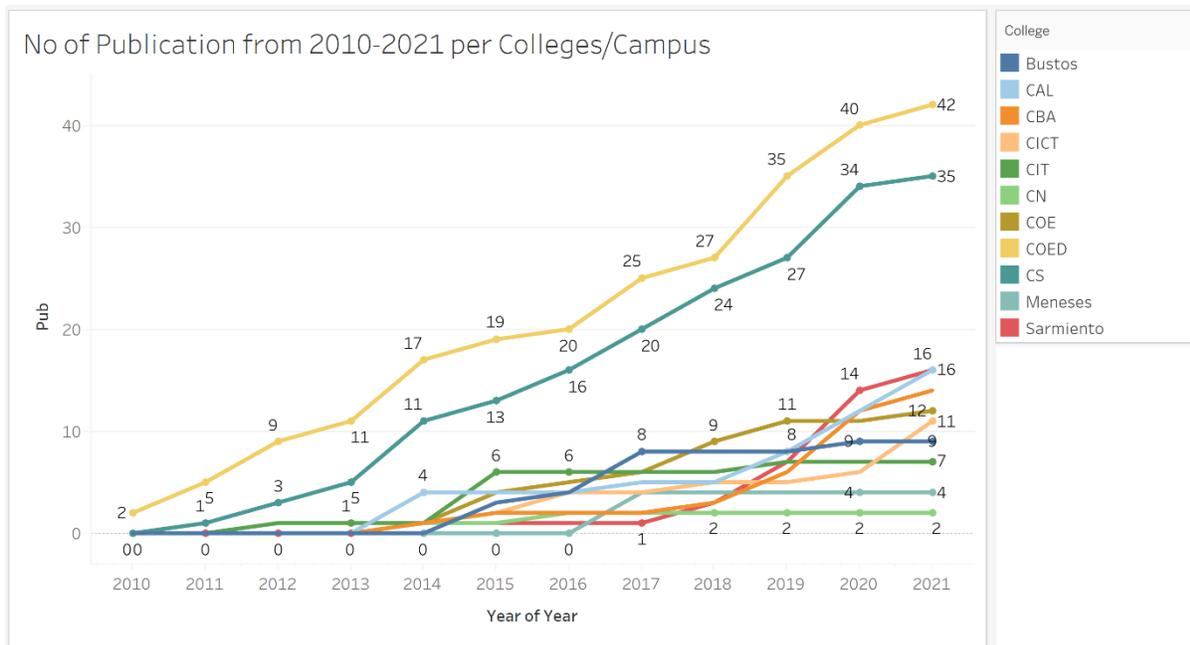

*Figure 4.* Number of Publication per Colleges and Campuses from 2010-2020

The reason behind the significant increase in the number of publications of the faculty members of BulSU-Sarmiento Campus is the scholarship grants and study break that the faculty members were able to receive which allowed them to have the luxury of time to produce quality researches and have them published in credible journals. As for the least number of publications produced, the College of Nursing (CN) has seen stagnation from the year 2018 to 2021 and were only able to publish two papers, and the same goes with the Meneses Campus which has seen no increase in the number of publications from the year 2014 to 2021. The result shows that some colleges and campuses have shown increase and progress in their publications but there are still some colleges and campuses that are yet to show progress in their publications.

Figure 5 shows that the College of Education (COEd) which has the leading number of publications has contributed 24% of the total number of publications found in Google Scholar. CS who is the next leading college has successfully contributed 21% to the total number of publications which shows a substantial amount of contribution. COE is the third leading college with a 12% contribution to the equivalent of the total

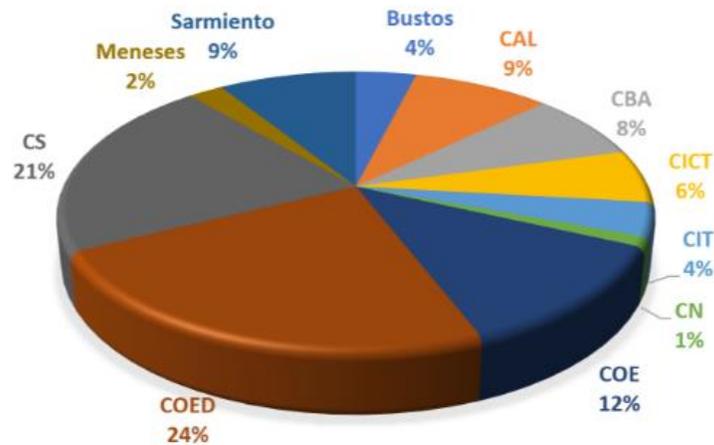

*Figure 5*. Percentage share per Colleges and Campuses in Publication

Other colleges like CAL, CBA, CICT, CBA, and CN have contributed less than 10% to the totality of the publications which means that they only have a few published types of research found in Google Scholar. The external campuses, Sarmiento, Bustos, and Meneses, have contributed a total of 15% in the totality of the publications under Bulacan State University in the Google Scholar database.

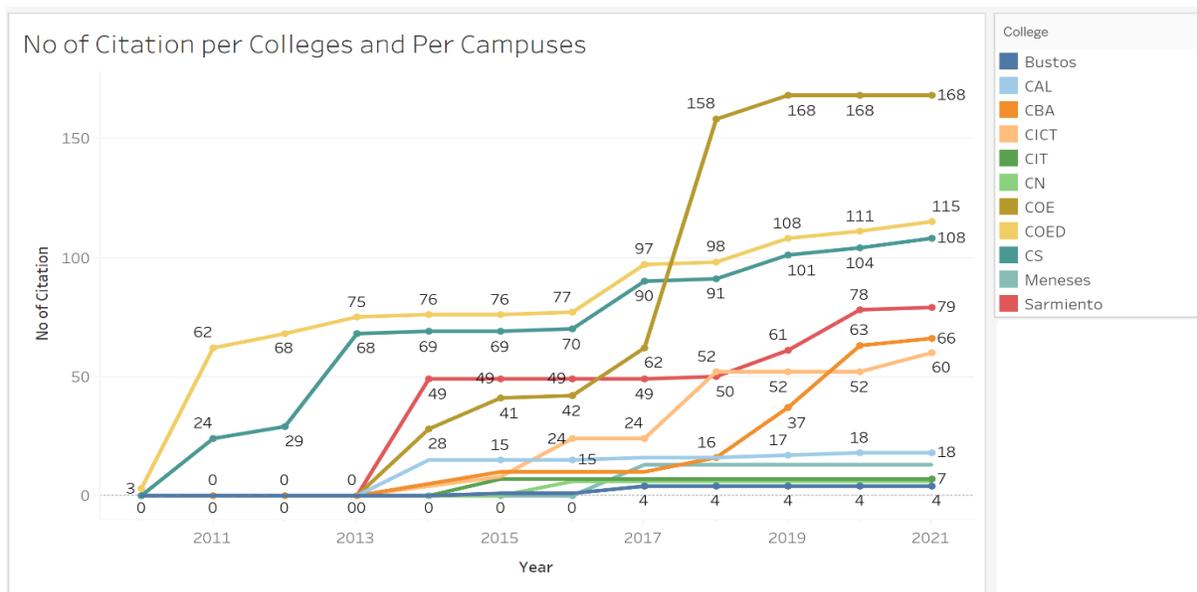

*Figure 6*. Insights on Journal Prestige and Citation Impact

The Cole Brothers who are two of the pioneers of citation studies describe citations as a measure of quality. A slightly more circumspect definition was also given in the introduction of their book on social stratification in science: "The number of citations is taken to represent the relative scientific significance or 'quality of papers' (Cole, 1992). With this, we can say that the number of publications a faculty member will not guarantee the number of citations that they can get as citations signifies the relevance and quality of the paper. In Figure 6 shown, the College of Engineering (COE) is easily the college that accumulated the highest number of citations. COE also showed a significant increase in the number of citations during the year 2017 – 2019 which was brought by the collaboration of the faculty members with other researchers from different universities and colleges. The collaborations allowed the possibility of the papers being exposed and introduced to different institutions and being recognized by a greater number of researchers. COEd and CS who were the top two colleges in BulSU in the number of publications have been steadily getting citations as well which proves that they are not just only publishing papers for the sake of adding numbers under their names but they made sure to publish papers that are of high quality and great relevance. Bustos Campus, on the other hand, has accumulated a lesser number of citations than CON and Meneses Campus despite having more publications than the two mentioned colleges.

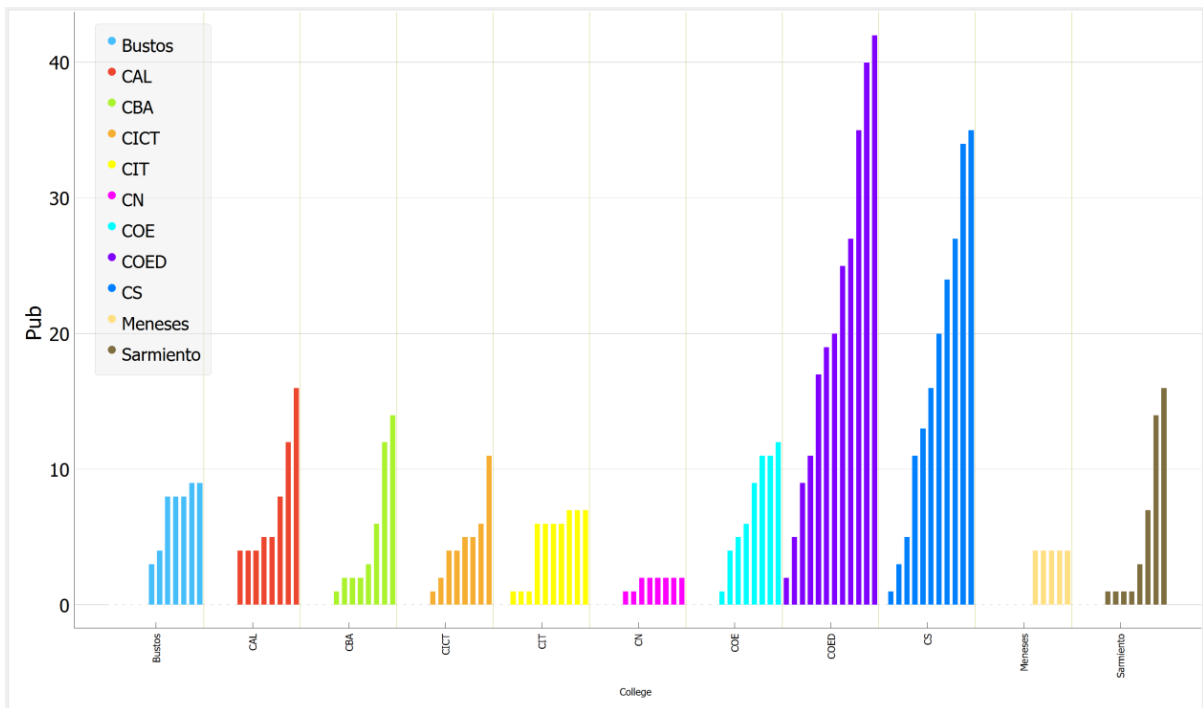

*Figure 7*. insights on Journal Publication per colleges from 2010-2021

The progress on the number of papers being published by the different colleges and campuses of Bulacan State University is clearly shown in Figure 7. the graph where COEd and CS have consistently boasted in publishing papers. As seen on the graph, the two colleges were able to produce more and more publications as the year progresses suggesting a strong research environment in the colleges. As for the other colleges and campuses like Sarmiento

Campus, CAL, and CBA, they showed a significant increase in the number of publications produced which proves the improvement of the faculty members in producing papers. On the contrary, as the other colleges and campuses continue to show a significant increase and steady progress in publishing their papers, there are still other colleges and campuses like Bustos Campus, CIT, CN, and Meneses Campus that recorded little to no progress in publishing papers.

Citations are often seen as an achievement of the research and the researchers themselves and this graph shows in Figure 8 the breadth of the citations received by the researchers of the BulSU faculty members from different colleges and campuses. The highest number of citations received goes to COE where faculty members are actively collaborating with researchers in the academe outside of Bulacan State University. The significant increase in numbers is very evident during the latter years of the covered years of this paper. COEd and CS have also shown growth in the number of citations received by the papers of the faculty members under those colleges. Bustos Campus, CIT, CN, and Meneses Campus have not seen any progress for years.

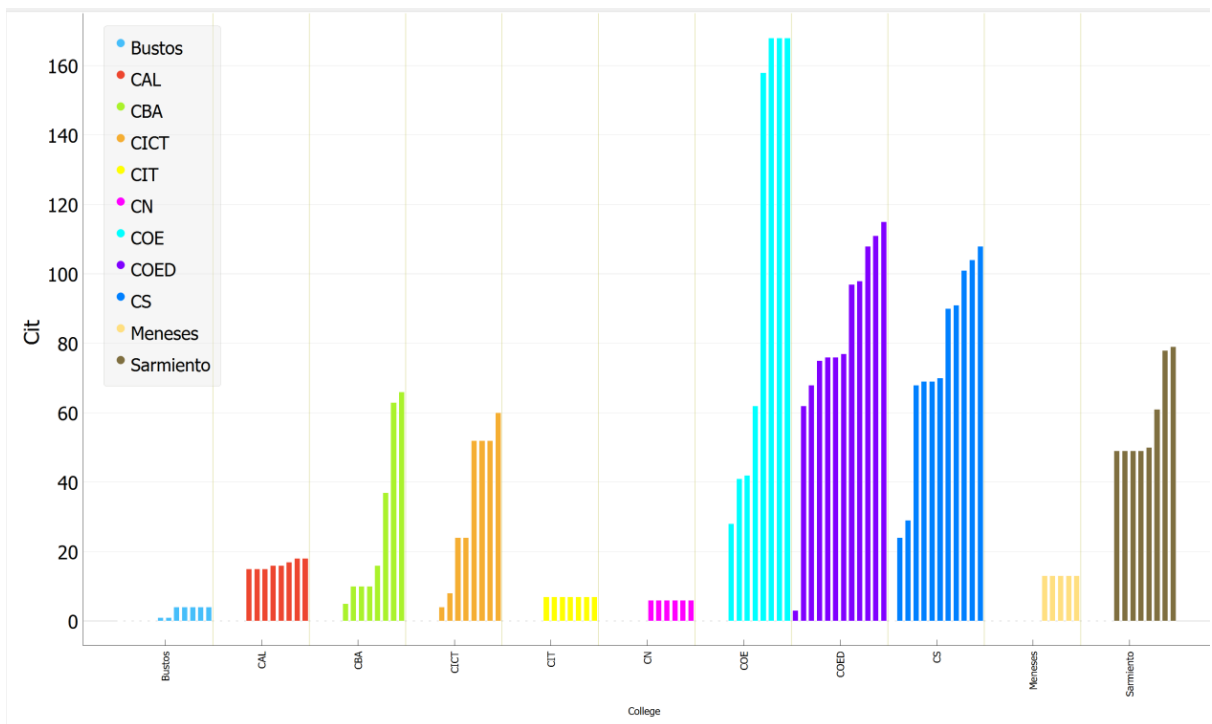

*Figure 8.* insights on Citation per colleges from 2010-2021

## CONCLUSION

Bulacan State University has seen both big improvements and stagnation with the growth of the number of publications being produced by the faculty members. It shows that BulSU continues to improve its Research and Development productivity. Based on the results of the study, the publications of BulSU have experienced a significant increase due to the impact of scholarship and the benefits that comes with it to the productivity of the faculty members. With the learnings that the scholarship offers and the luxury of time that the study break gives, the faculty members can produce more researches and are capable of publishing more papers. On the other hand, the factor that affected the significant increase in citations is the collaborators with researchers. With the opportunity of sharing perspectives with educators from other institutions, the chance of exposure and a greater depth for the research gets higher.

## RECOMMENDATION

Analyzing the number of publications and citations that the faculty members of Bulacan State University produce allows the monitoring of the productivity of the faculty members in terms of research and development. With putting the results and limitations of the study in mind, the administration should encourage the faculty members that engage in research activities to create Google Scholar accounts for easy monitoring of publications and citations. The researchers are aware of the limitations and the fact that not every faculty researcher was able to transfer their documents from physical publications to digital publications that's why it will be beneficial to both the administration and the faculty researcher if they create an account for easier monitoring. To tackle the stagnation that some of the campuses and colleges faced, aside from the encouragement to be part of the Google Scholar database, proper and additional incentives can be a positive reinforcement for the faculty researchers to experience once they've shown diligence in producing quality researches that attract multiple citations from other researchers. Collaborations with funding agencies such as DOST, DENR, JICA USAID would also be of great help and benefit to the faculty researchers especially when they plan to conduct big-scale research. Funding agencies are not only known for giving financial support but also capacity-building training that would capacitate BulSU's faculty members and improve the quality of the papers being produced for a higher chance of those papers to be published in highly-acclaimed journals locally and internationally.

## ACKNOWLEDGEMENT

The researchers extend their gratitude to the Bulacan State University Research Management Office-RMO for funding the research making the process of producing the paper possible. Appreciation and gratitude are also sent to the Bulacan State University Administration for always showing support and enthusiasm to assist faculty members who would like to get involved in research activities thus creating a healthy and encouraging research environment.